# Spatiotemporal Characteristics and Factor Analysis of SARS-CoV-2 Infections among Healthcare Workers in Wuhan, China

Peixiao Wang [a], Hui Ren [a], Xinyan Zhu [a,d,e,*], Xiaokang Fu [a], Hongqiang Liu [c] and Tao Hu [b*]

[a] State Key Laboratory of Information Engineering in Surveying, Mapping and Remote Sensing, Wuhan University, Wuhan 430079, China; peixiaowang@whu.edu.cn

[b] Center for Geographic Analysis, Harvard University, Cambridge, MA 02138, USA

[c] College of Geodesy and Geomatics, Shandong University of Science and Technology, Qingdao 266590, China; liuhongqiang@whu.edu.cn

[d] Collaborative Innovation Center of Geospatial Technology, Wuhan 430079, China

[e] Key Laboratory of Aerospace Information Security and Trusted Computing, Ministry of Education, Wuhan University, Wuhan 430079, China

[*] Correspondence: Xinyan Zhu (e-mail: xinyanzhu@whu.edu.cn) and Tao Hu (e-mail: taohu@g.harvard.edu)

**Conflicts of Interest:** The authors declare no conflict of interest.

**Acknowledgements:** This project was supported by the Key Program of National Natural Science Foundation of China (No. 41830645), National Key Research and Development Program of China (Grant Nos. 2018YFB0505500,2018YFB0505503), Funding program: CAE Advisory Project No. 2020-ZD-16.

# Spatiotemporal Characteristics and Factor Analysis of SARS-CoV-2 Infections among Healthcare Workers in Wuhan, China

**Summary:** Studying the spatiotemporal distribution of SARS-CoV-2 infections among healthcare workers (HCWs) can aid in protecting them from exposure. In this study, an open-source dataset of HCW diagnoses was provided, and the spatiotemporal distributions of SARS-CoV-2 infections among HCWs in Wuhan, China were explored. A geographical detector technique was then used to investigate the impacts of hospital level, type, distance from the infection source, and other external indicators of HCW infections. The results showed that the number of daily HCW infections over time in Wuhan followed a log-normal distribution, with its mean observed on 23$^{rd}$ January 2020 and a standard deviation of 10.8 days. The implementation of high-impact measures, such as the lockdown of the city, may have increased the probability of HCW infections in the short term, especially for those in the outer ring of Wuhan. The infection of HCWs Wuhan exhibited clear spatial heterogeneity. The number of HCW infections was higher in the central city and lower in the outer city. Moreover, HCW infections displayed significant spatial autocorrelation and dependence. Factor analysis revealed that hospital level and type had an even greater impact on HCW infections; third-class and general hospitals closer to infection sources were correlated with especially high risks of infection.

## 1. Introduction

At the end of 2019, SARS-CoV-2 was discovered and it spread rapidly around the world. By October of 2020, more than 200 countries had been infected, and there were more than 30 million confirmed cases and more than 900,000 deaths attributed to the virus [1]. At present, the management of and response to SARS-CoV-2 are common challenges facing all of humanity [2-4]. During the fight against the epidemic, healthcare workers (HCWs) have always been on the front line. While helping patients combat the virus, HCWs are also exposing themselves to high levels of the virus at close range [5, 6]. Thus, how to better protect HCWs is a key issue at both the city and national levels.

Since the outbreak of SARS-CoV-2 began, different degrees of HCW infections have occurred in many countries and regions [7, 8]. For example, the number of HCW infections in China account for ~4% of all infections nationwide [9, 10], while those in Italy account for ~10%. Studies have shown that the infection rate of HCWs is significantly higher than that of non-HCWs [11]. In order to deal with HCW infections, researchers in China and elsewhere have explored the causes of HCW infections [12], revealed the transmission routes leading to HCW infections [13], and formulated allocation schemes to address resource limitations [14] and protective measures [15, 16].

Spatiotemporal analysis involves the use of statistics, geographical and time-series data to study the patterns and mechanisms underlying a given phenomenon over time and space [17, 18]. In the field of public health, even the introduction of more simplified analytical methods to evaluate spatial and temporal relationships could be beneficial [19]. Thus, researchers used geographic information systems [20, 21] to understand the spatiotemporal patterns of viral transmission among reported cases in the early stages of the epidemic [22-24]. However, due to the lack of a data on HCW diagnoses, few existing HCW infection-related studies have been able to reveal the spatiotemporal characteristics of HCW infections and the external environmental factors influencing infections from a geographic perspective.

Therefore, we first studied the spatial distributions of HCW infections in Wuhan using partitioning statistics, distribution fitting, and spatial autocorrelation techniques [25, 26]. A geographical detector was then used to explore the impacts of external factors, such as hospital level, type, and distance from the infection source, on HCW infections [27]. Our findings can provide valuable insights into the spatiotemporal trends of SARS-CoV-2 infections among HCWs and the external factors influencing such infections, thus providing a foundation for improving the protection of HCWs. The primary contributions of this work include:

(1) A new method for generating infection inventories;
(2) Two HCW infection rate inventories—the Grid-Level Healthcare Worker Infection Inventory and the Hospital-Level Healthcare Worker Infection Inventory—based on HCW diagnoses in Wuhan;
(3) Clarification of the spatiotemporal characteristics and external influencing factors of HCW infections from a geographic perspective using the new HCW infection inventories;
(4) An open-source dataset of HCW diagnoses, which both ensures the reproducibility of the study and provides the data needed to support related research on HCW infections.

## 2. Methods

*2.1 Study Area and Data Sources*

Wuhan is not only the first city in China where the SARS-CoV-2 was discovered, but also the city with the most serious infections among HCWs in China. Since the outbreak began, more than 3,000 HCWs have been infected in Wuhan, accounting for 82.5% of infections nationwide. As shown in Figure 1, Wuhan is in Central China and is composed of a central city and an outer city. The central city includes the districts of Jiangan, Jianghan, Qiaokou, Hanyang, Wuchang, Qingshan and Hongshan, while the outer city includes Huangpi, Xinzhou, Dongxihu, Jiangxia, Caidian, and Hannan.

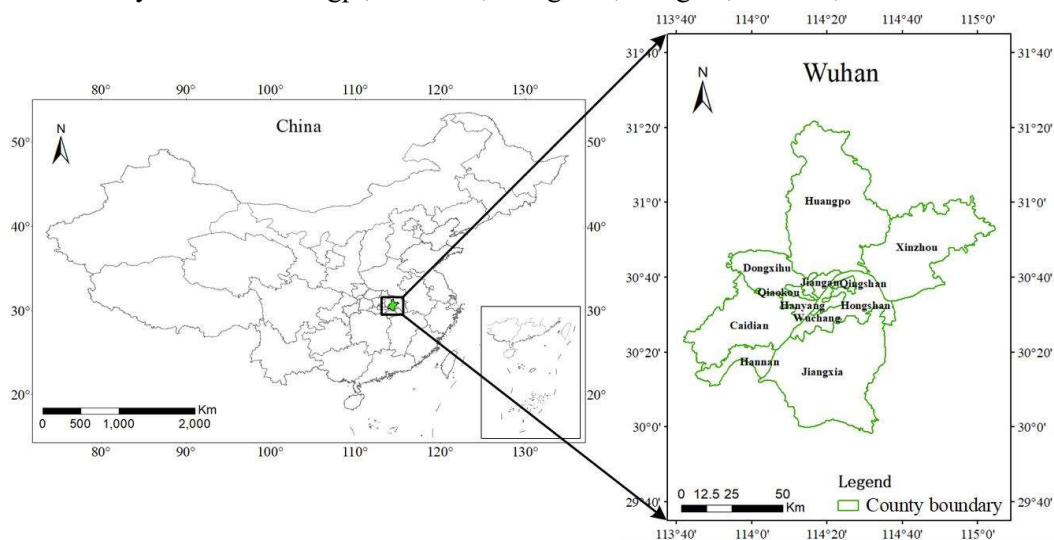

**Figure 1.** Map of the study area.

The data used in this study can mainly be divided into two types: (1) hospital data from Wuhan and (2) confirmed data on HCW infections in Wuhan.

Hospital data from Wuhan were mainly taken from the Health Commission of Hubei Province (http://wjw.hubei.gov.cn/). We used web crawler technology to search for medical institutions in Wuhan and parsed the returned data to obtain hospital information. After parsing, a total of 285 pieces of hospital information were obtained. Each record contained the unique identity, name, rank, type, latitude, and longitude of the hospital. Based on the hospital grading system of China [28], the ranks mainly include first-class, second-class, and third-class hospitals. The number of beds in first-class hospitals is generally less than 100, the number of beds in second-class hospitals is generally between 101 and 500, and the number of beds in third-class hospitals is generally greater than 501. The type mainly include community, specialised, and general hospitals. Community hospitals refer to medical institutions that provide basic medical services, general hospitals refer to medical institutions that have no restrictions on the scope of disease treatment, and specialist hospitals refer to medical institutions that have certain restrictions on the scope of disease treatment. To study the impact of the distance between a hospital and the infection source on HCW infections, we further calculated the straight-line distance between each hospital and the South China Seafood Market [29]; the calculated results are represented by the "distance" throughout this work.

Data on HCW diagnoses in Wuhan were mainly derived from the Chinese Red Cross Foundation (https://www.crcf.org.cn/), which distributes relief funds to every confirmed healthcare worker [30]. As of 11th September 2020, 83 groups of HCWs had received foundation assistance. We used crawler technology to obtain 3,703 publications that elucidated the conditions of HCWs that suffered from SARS-CoV-2, including 3,655 from Hubei Province and 3,058 from Wuhan City. Each record contains the province, city, and hospital name of confirmed HCW infections. In addition, the "hospital name" field of confirmed HCW infections is linked with the same field in the hospital data.

*2.2 Methodological Framework*

The methodological framework of this study is shown in Figure 2. First, two types of HCW infection inventories, based on the crawled HCW infection data, were constructed: the Grid-Level Healthcare Worker Infection Inventory and the Hospital-Level Healthcare Worker Infection Inventory. The infection inventory records daily infection number for HCWs at different levels. Using the Grid-Level Healthcare Worker Infection Inventory, we explored the spatiotemporal distributions and spatial autocorrelation of HCW infections. Finally, based on the Hospital-Level Healthcare Worker Infection Inventory, a geographical detector technique was used to explore the impacts of hospital level, type, distance from the infection source (i.e. South China Seafood Market), and other environmental indicators on HCW infections. The details of the methodology framework are described in appendices A.1–A.5.

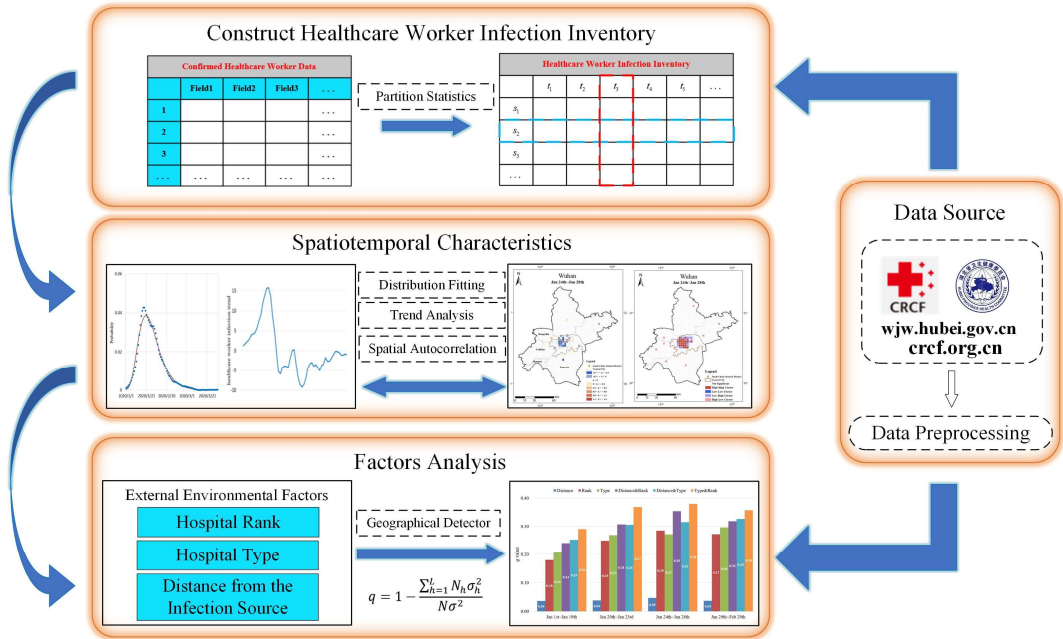

**Figure 2.** Research framework of spatiotemporal characteristics and correlation analyses for HCW infections.

## 3. Results and Discussion

### 3.1 Temporal Characteristics of HCW Infections

To assess the temporal characteristics of HCW infections, we first analysed the statistical distributions and changes in HCW infections over time. Figure S2 (see supplementary material) shows the results fitted to normal, log-normal, and gamma distributions for HCW infections in Wuhan. The mean square error (MSE) was used to evaluate the fit of each distribution. The results showed that a log-normal distribution was the most suitable to describe the evolution of HCW infections in Wuhan. Table S2 (see supplementary material) shows the statistical characteristics of HCW infections, for which 95% confidence intervals were calculated via bootstrap resampling. Based on the log-normal distribution, the mean day of infections in Wuhan was estimated to be 23$^{rd}$ January 2020, according to the data on daily cases of HCW infections, and the estimated standard deviation was 11.8 days.

We analysed the changes in the daily infection number of HCWs with time and the results are shown in Figure S3 (see supplementary material). Overall, HCW infections in Wuhan mainly occurred between 1$^{st}$ January and 29$^{th}$ February 2020. Within that period, infections rose from 1$^{st}$–19$^{th}$ January, while from 19$^{th}$ January to 29$^{th}$ February, they declined. Moreover, from 23$^{rd}$–28$^{th}$ January, the downward trend of HCW infections gradually decreased, and the infection rate curve showed an upward trend again between 25$^{th}$ January and 28$^{th}$ January, thereafter the HCW infections fell again. There are two main reasons for this change. First, the lockdown measures largely contained the spread of the SARS-CoV-2, but in a short period of time, it led to a large number of symptomatic patients visiting the hospitals, which increased the intensity of work for medical staff, leading to increased infection rates among them. Secondly, as of 28$^{th}$ January, there were more than 6,000 HCWs in Hubei, which greatly eased the workload, thus decreasing the rates of infection. Figure S1b clearly shows the changes in the trend of HCW infections over time. The decreasing trend of HCW infections began to slow on 21$^{st}$ January. This time node is highly consistent with that determined by human-to-human transmission.

In general, the temporal characteristics of HCW infections were highly correlated with the times of release of news or policies. In particular, the release of isolation measures (e.g. lockdowns) may exacerbate HCW infections in the short term. Therefore, before the implementation of such policies, it is particularly important to strengthen the measures in place to protect HCWs from infection.

*3.2 Spatial Characteristics of HCW Infections*

As shown in Figure S2, the infection of HCWs can be roughly divided into four stages: (1) 1st–19th January, (2) 20th–23rd January, (3) 24th–28th January, and (4) 29th January–29th February. Here, we show the spatial distributions and trends, and analyse the spatial autocorrelations during these four stages.

3.2.1 Spatial Distributions

The spatial distributions of HCW infections are shown in Figure S4 (see supplementary material). We found that the spatial distribution of HCW infections during the four stages showed the same trend, in that the rates of infection in the central urban area were higher than those in the outer urban area. This indicates that infections among HCWs in Wuhan were extremely unbalanced and spatially heterogeneous.

The areas with severe HCW infections were mainly located near the South China Seafood Market in the central urban area, as this market was a key source of the SARS-CoV-2 epidemic in Wuhan. However, there were slight differences in the spatial distributions of HCW infections between the four stages. Compared with the distribution of HCW infections from 1st–19th January, the rate of HCW infections in the outer city of Wuhan from 20th January to 29th February gradually increased.

3.2.2 Spatial Trends

The spatial trends of HCW infections are shown in Figure S5 (see supplementary material). From 1st–19th January, the overall rate of HCW infection in Wuhan increased rapidly, with the fastest growth rate in the areas near the South China Seafood Market. From 20th–23rd January, HCW infections exhibited a downward trend for the first time, among which the rate of decline was relatively fast in Wuchang and Jiangan. However, across Wuhan, the infection of HCWs continued to increase. From 24th–28th January, the rate of HCW infections in Wuhan showed a declining trend overall, while the rates of infection in Dongxihu, Caidian, Hannan, and Jiangxia showed an upward trend again. Combined with Figure S2, these results demonstrate that intense measures, such as the lockdown of Wuhan, may suddenly increase the workload of HCWs in mild and non-infected areas, and thus increase the probability of HCW infections in outer city areas. From 28th January to 29th February, the overall trend of HCW infection decreased, but there were still some areas with slight increases in rates of infection near the South China Seafood Market, indicating that HCWs near the infection source were always more susceptible to infection.

3.2.3 Spatial Autocorrelations

The Global Moran's I was used to explore the global spatial dependency of the four stages of HCW infections, as shown in Table S3 (see supplementary material). In each study interval, there was a significant spatial dependence of HCW infections, and with increases in time, this dependence gradually increased. From 24th–28th January, the spatial dependency was the strongest, where the Global Moran's I = 0.475, $Z > 2.58$, and $p < 0.001$, indicating that HCW infections in Wuhan were significantly and positively correlated at the grid-level. In other words, grids with high numbers of infections among HCWs had higher numbers of HCW infections across the grid, while grids with low numbers of HCW infections had lower numbers of HCW infections across the grid.

We used the Local Moran's I to explore the local spatial dependency of the four stages of HCW infections, as shown in Figure S6 (see supplementary material). Three clustering patterns were observed in the four stages, namely High-High, High-Low, and Low-High abnormal clustering. High-high clustering occurred in the infection hot spot area, which was mainly concentrated around the South China Seafood Market in the central city. High-low clustering was mainly distributed near the High-High clustering, indicating that HCW infections were essentially controlled within a certain range. Abnormal Low-High clustering was mainly distributed in the districts of the outer city in a discrete manner, and the Low-High areas gradually increased over time, indicating that some areas of the outer city also had a high incidence of HCW infections. Combined, the Global and Local Moran's I values reveal that HCW infections exhibited regional agglomeration.

*3.3 Analysis of Factors Related to HCW Infections*

In this study, we used geographic detectors to explore the impacts of external environmental factors, such as hospital level, type, and distance from the infection source, on HCW infections in the four stages previously identified. The results are shown in Figure S7 (see supplementary material). From a single-factor perspective, hospital level and type had greater impacts on HCW infections, with $q$-values mainly falling between 0.18 and 0.35, indicating that hospital type and rank could explain 18%–35% of HCW infections. Meanwhile, distance had a lesser impact on HCW infections, with $q$-values ranging from 0.03–0.04, indicating that distance could only explain 3% to 4% of HCW infections. To further verify the validity of the $q$ value, Table S4 showed the hypothesis test results of each variable. The results showed that the $p$ values of the three variables were all less than 0.05, that is, the three variables could explain the phenomenon of HCW infections. With respect to the interactions between two factors, hospital level and hospital type worked together to explain HCW infections to the greatest extent. At the same time, while the distance had a small impact on HCW infections, the combination of distance and either hospital level or type could cause a significant increase in the $q$-values, which indicates that, combined, these factors could explain HCW infections to a great degree. Additionally, with increases in time, the explanatory power of hospital level or type on HCW infections gradually increased, while the explanatory power of distance changed little.

To further explore the influences of hospital level, type, and distance from the infection source on HCW infections, the percentages of daily HCW infections to total HCW infections for each factor were computed (Table S5) (see supplementary material). The results show that HCWs in third-class and general hospitals were more likely to be infected. While distance had little explanatory power, HCWs who were close to the infection source were still more vulnerable to infection. As shown in Figure S6, HCWs at third-class and general hospitals closer to the infection source were at the greatest risk of infection.

## 4. Conclusions

As the first line of defence in the fight against the SARS-CoV-2 epidemic, HCWs help patients combat the virus, while exposing themselves to high concentrations of the virus. It is important to study the spatiotemporal distributions and factors influencing SARS-CoV-2 infections among HCWs to better protect them. Existing studies related to HCW infections have largely emphasised infection rates and protective measures, while the spatiotemporal patterns and related external factors related to HCW infections have been unclear. To fill this gap in the existing knowledge, we first studied the spatiotemporal distributions of infections of HCWs in Wuhan using statistical partitioning, curve-fitting, and spatial autocorrelation. The results showed that the lowest MSE was obtained using a log-normal distribution, which indicates that this distribution was the most suitable to describe the evolution of HCW infections over time. The mean date of infection was 23rd January, with a standard deviation of 10.8 days. The implementation of substantial measures, such as the lockdown of Wuhan, may have contributed to short term increases in the probability of infection among HCWs, especially in outer city areas. This indicates that it is especially important to strengthen the protective measures for HCWs before such policies are implemented. The rate at which HCWs were infected in Wuhan displayed apparent spatial heterogeneity. The number of HCW infections was higher in the central city and lower in the outer city and the likelihood of infection exhibited significant spatial autocorrelation and spatial dependency.

To better understand the temporal and spatial distributions of HCW infections, geographical detector techniques were used to explore the impacts of external environmental factors, such as hospital level, type, and distance from the infection source on HCW infections. The results showed that hospital level and type significantly impacted the probability of HCW infections, among which third-class and general hospitals carried the greater risks of infection. This conclusion is similar to that of Zheng et al. [11], which independently validates the accuracy of our results to a certain extent. Moreover, although the distance from the infection source had little impact on HCW infections, the combined effect of distance and hospital level or type was shown to increase the explanatory power of HCW infections, that is, third-class and general hospitals closer to the infection source had the greatest risks of HCW infection. Therefore, investing more medical supplies into hospitals with higher risks of HCW infection is vital.

This study had four primary limitations. First, the selection of influencing factors excluded factors that are known to contribute to the incidence of HCW infections, such as access to sufficient medical supplies, sufficient knowledge of protective protocols, and patient contact, among others. However, due to the difficulty of data acquisition, only three external factors — hospital rank, type, and distance from the infection source — were considered. Secondly, the coverage of HCW infection data was narrow. Only data on HCW infections in China were used to explore related spatiotemporal characteristics and influencing factors; data from other countries were not included, which may limit the generalizability of our results. Third, the infection data for HCWs was not comprehensive. As the information on HCW infections is published on the official website of the Chinese Red Cross Foundation in batches, additional information may continue to be published in the future that could not be included here. Finally, we used the number of HCW infections instead of the infection rate, ignoring the heterogeneity of HCWs in Wuhan. In light to these limitations, future studies should focus on collecting additional domestic and foreign HCW infection data to more accurately and comprehensively analyse the spatiotemporal characteristics and factors influencing SARS-CoV-2 infections among HCWs.